\documentclass{article}
\usepackage{fortschritte,graphicx}
\def\np{Nucl. Phys.}
\def\pl{Phys. Lett.}
\def\prl{Phys. Rev. Lett.}
\def\pr{Phys. Rev.}

\def\ijmp{Int. J. Mod. Phys.}

\def\jhep{J. High Energy Phys.}

\def\rc{\nonumber\\}
\def\ZZ{{\rm Z}\kern-3.8pt {\rm Z} \kern2pt}

\def\beq{\begin{equation}}                     %
\def\eeq{\end{equation}}                       %
\def\bea{\begin{eqnarray}}                     
\def\eea{\end{eqnarray}}                       
\begin {document}     
\begin{flushright} \vspace{-2cm} 
{\small  CPHT-PC 067.1204\\ 
hep-th/0412105} \end{flushright}            
\def\titleline{
%
%
%
Supersymmetry and flavor in a                    
sugra dual of ${\cal N}=1$ Yang-Mills                              
}
\def\email_speaker{
{\tt 
%
%
paredes@cpht.polytechnique.fr             
}}
\def\authors{
%
%
%
%
%
A. Paredes\1ad 
}
\def\addresses{
%
%
%
%
\1ad                                        %
Centre de Physique Th\'eorique, \'Ecole Polytechnique\\            %
91128, Palaiseau Cedex, France\\            
%
}
\def\abstracttext{
%
%
Some field theory aspects can be addressed holographically by
introducing D-branes in the string theory duals.
In this work, we study this issue in the framework of a dual of
${\cal N}=1$ YM, the so-called Maldacena-N\'u\~nez model.
Some supersymmetric embeddings for D5-brane probes are found and
they are interpreted as the addition of fundamental flavors to
the gauge theory. This allows to give a dual description of
some known aspects of ${\cal N}=1$ SQCD and to compute a mass
spectrum of mesons.
}
\large
\makefront
\section{Introduction}
                 
The duality between string theories and large $N_c$ gauge theories
proposed by 't Hooft has led to surprisingly fruitful 
results during the last years. The best understood example
relates type IIB string theory on $AdS_5 \times S^5$ to
${\cal N}=4$ super Yang-Mills in four dimensions.
 Nevertheless, a large amount 
of work has also been devoted to the extension of the duality to
less symmetric, more realistic, setups. The interest of this
generalization is obvious and the final goal would be to find a
string dual of large $N_c$ QCD.

In order to develop models with reduced supersymmetry, one possibility
is to consider branes wrapping supersymmetric cycles inside special
holonomy manifolds. In some cases, it is possible to construct the
corresponding supergravity solution.
In this note, we will deal with the 
Maldacena-N\'u\~nez (MN) model \cite{MN}, constructed along these lines.

D-branes are very useful to provide an insight of how some field
theory aspects can be recognized in the dual setup.
In particular,  fields transforming
in the fundamental representation of the gauge group must be dual to
open strings with just one end on the gauge theory branes. This suggests
that, in order  to add flavor to
a gauge theory living on a stack of $N_c$
D-branes, one should add another stack of $N_f$ D-branes. 
We will review the explicit realization 
 of this idea in the framework of the 
MN model \cite{flavoring}.

\section{The gravity solution and its Killing spinors}

The MN model was proposed as a string
theory dual of ${\cal N}$=1 $SU(N)$ Yang-Mills in the IR. The setup
consists of a stack of $N_c$ D5-branes extended in 1+3 Minkowski
dimensions and moreover wrapped on the finite two-cycle inside
a conifold. When one decouples the modes on the sphere, the open
strings attached to the branes give effectively a four dimensional
gauge theory. The conifold breaks the supersymmetry to 1/4
and the branes further half it, so there are four supercharges left
corresponding to one Majorana spinor in four dimensions.

The corresponding supergravity solution was computed in 
\cite{CV}. It realizes a geometric transition so the two-cycle
shrinks and a finite three-cycle supported by
RR flux is left.

The (non-singular) metric can be written in terms of the 
following vielbein
 (type IIB theory, Einstein frame):
\bea
&&e^{x^i}=e^{{\phi\over 4}}\,d x^i\,\,\,
\,(i=0,1,2,3)\,\,,\,\,\,\,\,
e^{1}=e^{{\phi\over 4}+h}\,d\theta\,\,,
\,\,\,\,\,\,\,\,\,\,\,\,\,\,
e^{2}=e^{{\phi\over 4}+h}\,\sin\theta d\varphi\,\,,\rc
&&e^{r}=e^{{\phi\over 4}}\,dr\,\,,
\,\,\,\,\,\,\,\,\,\,\,\,\,\,
e^{\hat i}={e^{{\phi\over 4}}\over 2}\,\,
(\,w^i\,-\,A^i\,)\,\,\,\,\,\,\,(i=1,2,3)\,\,,
\label{frame}
\eea
where:
\beq
A^1\,=\,-a(r) d\theta\,,
\,\,\,\,\,\,\,\,\,
A^2\,=\,a(r) \sin\theta d\varphi\,,
\,\,\,\,\,\,\,\,\,
A^3\,=\,- \cos\theta d\varphi\,.
\label{oneform}
\eeq
and the $w$'s are $SU(2)$ left invariant one-forms parameterizing
the $S^3$:
\bea
w^1&=& \cos\psi d\tilde\theta\,+\,\sin\psi\sin\tilde\theta
d\tilde\varphi\,\,,\,\,\,\,\,
w^2=-\sin\psi d\tilde\theta\,+\,\cos\psi\sin\tilde\theta 
d\tilde\varphi\,\,,\rc
w^3&=&d\psi\,+\,\cos\tilde\theta d\tilde\varphi\,\,.
\eea
The functions appearing above and the dilaton are, explicitly: 
\beq
a(r)={2r\over \sinh 2r}\,\,,\,\,\,\,\,\,\,\,\,\,\,\,
e^{2h}=r\coth 2r\,-\,{r^2\over \sinh^2 2r}\,-\,
{1\over 4}\,\,,
\,\,\,\,\,\,\,\,\,\,
e^{-2\phi}=e^{-2\phi_0}{2e^h\over \sinh 2r}\,\,.
\label{MNdilaton}
\eeq
Finally, the D5-branes are source for a RR three-form field strength:
\beq
F_{(3)}\,=\,-{1\over 4}\,\big(\,w^1-A^1\,\big)\wedge 
\big(\,w^2-A^2\,\big)\wedge \big(\,w^3-A^3\,\big)\,+\,{1\over 4}\,\,
\sum_a\,F^a\wedge \big(\,w^a-A^a\,\big)\,\,,
\label{RRthreeform}
\eeq
where $F^a$ is defined as an su(2) field strength: 
$F^a\,=\,dA^a\,+\,{1\over 2}\epsilon_{abc}\,A^b\wedge A^c\,\,.$
In the following, we will need the explicit expressions of
the Killing spinors of the solution \cite{flavoring}:
\beq
\epsilon\,=\,e^{{\alpha\over 2}\,\Gamma_1\hat\Gamma_1}\,\,
e^{{\phi\over 8}}\,\,\eta\,\,,
\label{spinor}
\eeq
where $\eta$ is a constant spinor satisfying:
\beq
\Gamma_{x^0\cdots x^3}\,\Gamma_{12}\,\eta\,=\,\eta\,\,,
\,\,\,\,\,\,\,\,\,\,\,\,
\Gamma_{12}\,\eta\,=\,\hat\Gamma_{12}\,\eta\,\,,
\,\,\,\,\,\,\,\,\,\,\,\,
\eta\,=\,i\eta^*\,\,.
\label{project}
\eeq
Indices are referred to the frame defined in eq. (\ref{frame}).
On the other hand, $\alpha$ is a function of the radial variable
which introduces the necessary rotation on the Killing
spinor \cite{Twist} and reads:
\beq
\cos\alpha\,=\,{\rm \coth} 2r\,-\,{2r\over \sinh^22r}\,\,.
\label{alphaexplicit}
\eeq
Reviews of the model and its relation with the dual gauge theory
can be found in \cite{reviewMN}.

\section{Supersymmetric brane probes}

Different gauge theory objects are described holographically
by the introduction of new branes in the setup (see, for
instance, \cite{LoewySonn}). In this setup, D5-branes wrapping
the $S^3$ are domain walls, fundamental strings represent 
QCD strings, D3-branes wrapping an $S^2$ inside the $S^3$ are
bound states of fundamental strings and D3-branes wrapping
the $S^3$ give the baryon vertex. By using
(\ref{spinor}), (\ref{project}), one can check that the supersymmetry
preserved agrees with what is expected from the field theory.

The goal of this work is to explain the introduction 
in the theory of dynamical quarks. 
This can be achieved by
introducing branes extended in all Minkowski directions.
They should also stretch to infinity in the holographic (radial)
direction. Then, their 
infinite volume makes the open strings with both ends on
these new branes decouple from the theory. The open strings
with one end on each kind of branes are fundamental matter in
the gauge theory \cite{KK}. This reasoning has also been followed to
add flavor to the theories on the conifold \cite{Ouyang} and also 
non supersymmetric theories \cite{Johana}.

In the Maldacena-N\'u\~nez model, one has to consider D5-branes
extended ortogonally to the field theory branes inside the conifold
\cite{WangHu}. In principle, one should find a new gravity 
solution which includes the backreaction of these new branes.
However, finding this kind of backreacted solutions
is in general very difficult (for recent progress in this topic
in different scenarios, see \cite{backreacted}).
Therefore, we will treat the flavor branes as probes living in
the background. 
This, of course, limits the analysis to the quenched approximation
$N_f \ll N_c$.
The physics of the probes is described by the
Born-Infeld + Wess-Zumino action, and one can study their
supersymmetry using the kappa-symmetry techniques.
As we know from the field theory side that quark multiplets can be 
introduced without further breaking supersymmetry (obtaining
${\cal N}=1$ SQCD), this will be our guiding principle.

We will  require the fulfilment of:
\beq
\Gamma_{\kappa}\,\epsilon\,=\,\epsilon\,\,,
\label{kappaprojection}
\eeq
without introducing projections on the Killing spinor different from
(\ref{project}). This yields a system of first order equations for the
functions defining the
possible embeddings of the probes.

\section{Holographic flavor}
Equation (\ref{kappaprojection}) for a D5-brane probe
leads, in general, to a very
complicated system of equations. This problem was addressed in
\cite{flavoring} where several solutions were found. 

\begin{figure}[h]
\centerline{\includegraphics[width=15cm]{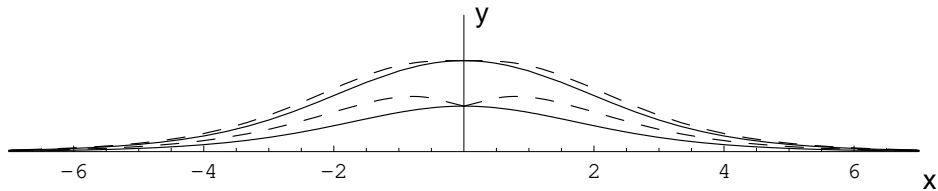}}
\caption{Pictorial representation comparing the embeddings defined
in (\ref{abflavor}) (dashed line) and (\ref{noabflavor}) (solid line)
 for the same value of $r_*$. The variables on each axis are
 $x=r \cos\theta$, $y=r \sin\theta$.
   The curves for two different values of $r_*$
($r_*=0.5$ and $r_*=1$) are shown. The massless
($r_*=0$) limit would be a brane extended along the $x$ axis at
$y=0$.}
\label{fig2}
\end{figure}

Let us focus on the solutions more interesting from the field theory
point of view,  those that can be naturally interpreted as
flavor branes. These embeddings are the equivalent of the 
hypersurfaces considered
in \cite{CIV}.
 Let us first consider the far UV of the gauge theory.
In the gravity solution, it is described by neglecting terms exponentially
small in $r$, {\it i.e.} taking $a(r)=0$. Then a solution 
of (\ref{kappaprojection}) is found by identifying the two-sphere
and a two-sphere inside the three-sphere:
\beq
\tilde\theta=\theta\,\,,
\,\,\,\,\,\,\,\,\,\,\,\,\,\,\,\,\,\,\,
\tilde\varphi\,=\,\varphi+\varphi_0,
\label{unitwin}
\eeq
where $\varphi_0$ is an arbitrary integration 
constant\footnote{Another possibility is to take
$\tilde\theta=\pi-\theta$, $\tilde\varphi\,=\,2\pi-\varphi+\varphi_0$\ .}.
Moreover, one has to set:
\beq
\psi=\psi_0
\,\,,\,\,\,\,\,\,\,\,\,\,
e^r\,=\,{e^{r_*}\over \sin\theta}\,\,\,\,
\qquad\ (a(r)=0)\,\,\,\,,
\label{abflavor}
\eeq
$\psi_0$, $r_*$ being two new integration constants. This embedding
can be generalized when considering the full gravity solution and
therefore exploring the IR of the gauge theory. Eq. (\ref{unitwin})
remains unchanged, but instead of (\ref{abflavor}),
one finds:
\beq
\psi=\pi,3\pi
\,\,,\,\,\,\,\,\,\,\,\,\,
\sinh r\,=\,{\sinh r_*\over \sin\theta}\,\,\,.
\label{noabflavor}
\eeq
These embeddings are depicted in figure \ref{fig2}.
Let us comment on some gauge theory aspects that can be read from
the above configurations. First of all, let us analyze the global
symmetries of the theory. The geometric dual of the R-symmetry of
the theory is related to shifts of the $\psi$ angle. It has been shown that
the pattern of symmetry breaking of the unflavored theory is explicitly
realized by the gravity solution \cite{MN}. 
The spontaneous breaking to $\ZZ_2$
in the IR is related to the formation of a gaugino condensate $<\lambda^2>$,
whose gravitational counterpart is the function $a(r)$. In the theory 
with flavor, there is also a squark condensate $<\Phi\bar\Phi>$ which
realizes this breaking. Therefore, it is appealing to find this behavior
from the probe point of view, as in the IR (\ref{noabflavor}) two
particular values of the constant $\psi_0$ are selected. There is
also a $U(1)$ symmetry unbroken by the solution, related to
$\varphi_0$. This can be identified with the baryonic $U(1)$ of the gauge
theory which stays unbroken, in agreement with a theorem by
Vafa and Witten \cite{VafaWitten}.

The constant of integration $r_*$ can be identified with the mass of the
quarks, as it gives the minimal distance between the gauge theory branes
and the flavor branes and therefore determines the 
minimal energy of an open string
stretching between them. This allows us to consider the massless limit
by taking $r_* \to 0$ in (\ref{noabflavor}). Clearly, now the branes
are extended in $r$ at fixed $\theta=\tilde\theta=0,\pi$. But this limit
is not continuous as the brane breaks into two pieces and because, moreover,
one cannot consider fluctuations of the type described below. This is the
holographic counterpart of the fact that, for massless flavors, there is
a runaway potential in the field theory that does not allow the formation
of $<\lambda^2>$, $<\Phi\bar\Phi>$, and therefore there is no stable vacuum. 
 Moreover, one
can check that in this limit one can again take any $\psi=\psi_0$ as expected
because there are no condensates that break the R-symmetry.

\begin{figure}[h]
\includegraphics[height=5cm]{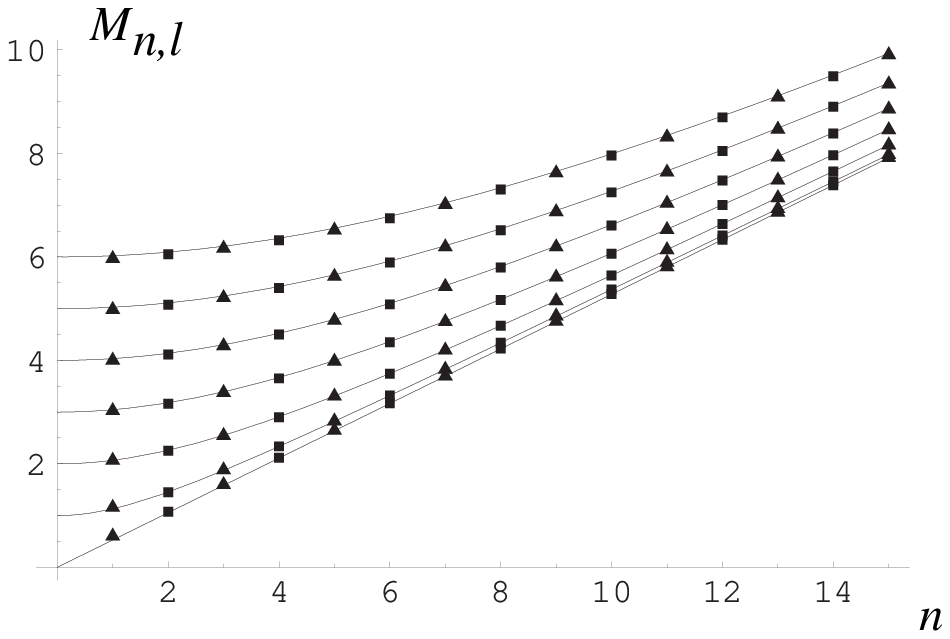}
\includegraphics[height=5cm]{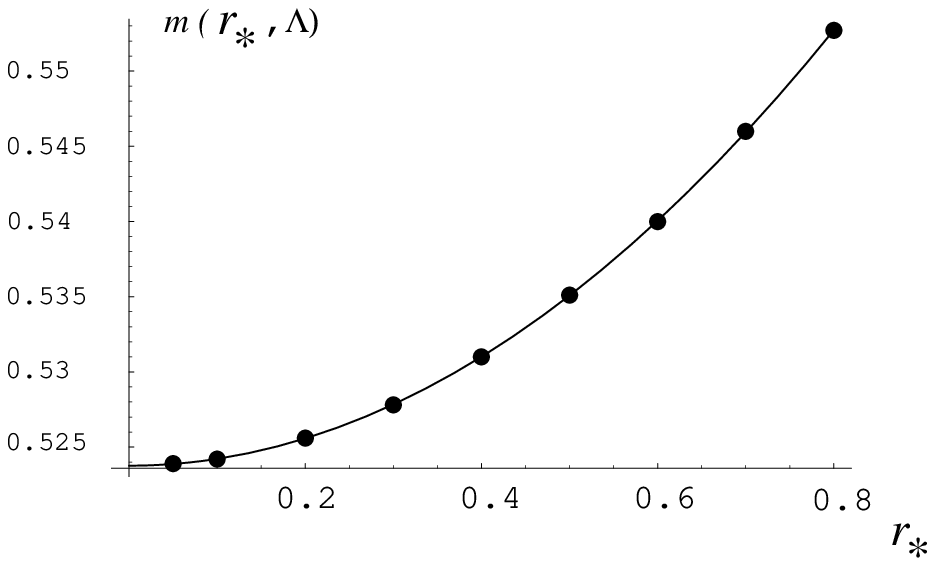}
\caption{ The spectrum of mesons found numerically. The figure on
the left gives the tower of 
meson masses in terms of the quantum number $n$ for
different values of $l$ ($r_*=.3$, 
$\Lambda=3$). The one on the right depicts 
(for $\Lambda=3$) the lightest
meson mass in terms of $r_*$, which can be identified with
the mass of the dynamical quarks. The solid lines are given by the
equations (\ref{spectrum}), (\ref{parabola}) respectively.
}
\label{fig}
\end{figure}

The spectrum of low energy excitations of the brane probes describes
the physics of the open strings (quarks), and, therefore, one can
identify it with the low energy states of the gauge theory, {\it i.e.}
the mesons \cite{KMMW}. With this purpose, let us consider
quadratic fluctuations around the stable embedding:
\beq
r(\theta, x,\varphi)\,=\,r_0(\theta)\,+\,
e^{ikx}\,e^{il\varphi}\,\,\zeta (\theta)\,\,,
\eeq
where $r_0(\theta)$ corresponds to the solution (\ref{noabflavor}). By 
inserting this ansatz in the Born-Infeld action, one finds the lagrangian
determining the dynamics of the system. Then, one would like to
compute the normalizable modes which should give the quantized, physical,
oscillations and get an expression for the allowed masses $M^2=-k^2$.
Unfortunately, this is not possible because all modes have infinite norm.
This is related to the fact that the solution cannot be trusted for
large $r$ as the dilaton grows unbounded. The same problem was found
when computing glueball masses in this background \cite{Pons}.
One can avoid this problem by introducing an UV cutoff $r=\Lambda$ for
the fluctuations. This is not unnatural since mesons are an IR effect.
The cutoff should not introduce a new scale in the setup. One should
take the scale separating the IR from the UV, which coincides with the
scale at which the dilaton becomes large. A precise value 
of $\Lambda$ cannot be given,
but it can be estimated to be around $\Lambda\approx 3$.
 One may think about this
procedure as taking the difference between the oscillations around
(\ref{abflavor}) and (\ref{noabflavor}) which are almost identical
at scales above $\Lambda$. Analyzing the equations
of motions by means of numerical computation, we found that the
mass spectrum for the scalar mesons
 is compatible with:
\beq
M_{n,l}(r_*,\Lambda)\,=\,\sqrt{m^2(r_*,\Lambda)\,\,n^2+l^2}
\label{spectrum}
\eeq
where $m(r_*,\Lambda)$ stands for the mass of the lightest meson. 
The quantum number $l$ comes from the KK modes not present in the
dual field theory and so it should be set to zero for the physical case.
Finally, one can also find numerically the dependence of the meson
masses on $r_*$, {\it i.e.} the quark mass.
\beq
m(r_*,\Lambda)\,=\,{\pi\over 2\Lambda}\,+\,b(\Lambda)\,r^{2}_*\,\,.
\label{parabola}
\eeq
Our numerical results compared to the models (\ref{spectrum}),
(\ref{parabola}) are depicted in fig. \ref{fig}.

It is also worth to point out that it is also possible to deal with
vector mesons by considering excitations of the gauge field living
on the brane probes. The relevant ansatz is:
\beq
{\cal A}_{\mu} (\theta,x,\varphi)\,=\,\epsilon_{\mu}\varsigma (\theta)\,\,
e^{ikx}\,\,e^{il\varphi}\,\,,
\eeq
where $\epsilon_{\mu}$ is a constant polarization four-vector.
The numerical computation yields an spectrum almost identical to
the one computed for scalar mesons (\ref{spectrum}), (\ref{parabola})
and therefore the model predicts a degeneracy between both kinds of states.
This fact should be related to the supersymmetry of 
the theory living on the branes.

\

{\bf Acknowledgement} It is a pleasure to thank C. N\'u\~nez and
A.V. Ramallo for collaboration on these topics.
I would also like to thank the organizers of the $37^{\textrm{th}}$
Symposium Ahrenshoop for their kind invitation. This work 
was partially supported by RTN contracts 
MRTN-CT-2004-005104 and
MRTN-CT-2004-503369 and by a European Union Excellence Grant,
MEXT-CT-2003-509661.


\end{document}